\documentclass[aps,prl,twocolumn,groupedaddress,showpacs]{revtex4}
\usepackage{graphicx,amsmath,amssymb}
\usepackage{bm}
\begin{document}
\title{Eulerian and Lagrangian velocity statistics in weakly forced
  two-dimensional turbulence} \author{Michael K. Rivera}
\affiliation{The Condensed Matter and Thermal Physics Group (MPA-10)
  and The Center for NonLinear Studies (T-CNLS), Los Alamos National
  Laboratory, Los Alamos, NM, 87545} \author{Robert E. Ecke}
\affiliation{The Condensed Matter and Thermal Physics Group (MPA-10)
  and The Center for NonLinear Studies (T-CNLS), Los Alamos National
  Laboratory, Los Alamos, NM, 87545} \date{\today}
\begin{abstract}
  We present statistics of velocity fluctuations in both the
  Lagrangian and Eulerian frame for weakly driven two-dimensional
  turbulence.  We find that simultaneous inverse energy and enstrophy
  ranges present in the Lagrangian and Eulerian Fourier spectra are
  not directly echoed in real-space moments of velocity difference.
  The spectral ranges, however, do line up very well with ratios of
  the real-space moments {\em local} exponents, indicating that though
  the real-space moments are not scaling ``nicely'', the relative
  behavior of the velocity difference probability distribution
  functions is changing over very short ranges of length scales.
  Utilizing this technique we show that the ratios of the local
  exponents for Eulerian moments in weak two-dimensional turbulence
  behave in agreement with Kolmogorov predictions over the
  spectrally identified ranges.  The Lagrangian local exponent ratios,
  however, behave in a different manner compared to their
  Eulerian counterparts, and deviate significantly from what would be
  expected from Kolmogorov predictions.
\end{abstract}
\pacs{abc.123}

\maketitle

\section{Introduction\label{sec: Introduction}}

There are two reference frames that are normally considered in
turbulent fluids: the Eulerian frame and the Lagrangian frame
\cite{Batchelor-Book-1982,Frisch-Book-1995}.  The Eulerian frame of
motion is fixed to the laboratory frame where velocities, pressures
and accelerations are fields fixed in space and varying with time
({\em i.e.,} the velocity field ${\bf u}({\bf x},t)$).  This frame has
been used for many classical approaches to the turbulence problem,
including the derivation of the Karman-Howarth equation and Kolmogorov
scaling theory.  The Lagrangian frame is fixed to fluid elements as
they are advected by the turbulence.  The position of the Lagrangian
fluid element in time is denoted ${\bf x}(t)$, and the velocity and
acceleration of the element are the first and second time derivative
of the Lagrangian trajectory, respectively.  The Lagrangian velocity
at any time for any trajectory must equal its Eulerian counterpart
({\em i.e.,} ${\bf u}(t) = {\bf u}({\bf x}(t),t)$).  The Lagrangian
frame is a useful one when considering the mixing of scalars
\cite{Monin-Book-1975,Pope-Book-2000,Falkovich-RMP-2001}, such as a
dye, by turbulent motion. The difficulties in making experimental
measurement of Lagrangian trajectories has made the characterization
of Lagrangian turbulence much less common, with a few results reported
for two-dimensional turbulence \cite{Jullien-PRL-1999,Rivera-PRL-2005}
and for three-dimensional turbulent flows
\cite{Ott-JFM-2000,Xu-PRL-2006}.

In either the Eulerian or Lagrangian frame, we would like to
characterize the turbulent state by studying the statistical nature of
the velocity fields by measuring, for example, spectra and moments of
velocity differences. We define the Eulerian $n^{\rm th}$-order moment
of longitudinal velocity difference as
\begin{equation}
S^{(n)}(r) \equiv \langle ( ({\bf u}({\bf x}+{\bf r}) - {\bf u}(\bf x) ) \cdot {\bf r})^{n} \rangle,
\end{equation}
and the Lagrangian $n^{\rm th}$ order moment of velocity difference as
\begin{equation}
D^{(n)}(\tau) \equiv \langle  |{\bf u}(t+\tau) - {\bf u}(t)|^{n}  \rangle.
\end{equation}
We would like to extract ranges of spatial scales $r$ for the
Eulerian frame and time scales $\tau$ for the Lagrangian frame over
which the velocity statistics exhibit scaling and measure the corresponding
scaling exponents. These exponents are then compared with theoretical
prediction, when such predictions exist.

The moment characterization of the velocity fluctuations of turbulence
is a standard procedure in fluid turbulence.  Here we follow Frisch
\cite{Frisch-Book-1995} with respect to this standard analysis
paradigm.  Velocity fluctuation moments have been experimentally
obtained in three-dimensions for both the Eulerian and Lagrangian
frames \cite{Sreenivasan-PTP-1998,Xu-PRL-2006}.  Kolmogorov proved
that for three-dimensional homogenous isotropic turbulence,
$S^{(3)}(r) = -4/5 \epsilon r$, where $\epsilon$ is the energy
dissipation rate.  Using dimensional analysis and the hypothesis of
strict self-similarity, one can then show that $S^{(n)}(r) \propto
(\epsilon r)^{n/3}$.  Subsequent measurements of fluctuations in
three-dimensional turbulent fluids demonstrate that strict
self-similarity does not hold in these systems, and that there are
significant deviations from the expected $n/3$ scaling exponent,
especially as $r$ approaches the viscous dissipation scale and $n$
becomes large \cite{Frisch-Book-1995}.  This deviation is attributed
to ``intermittency'', a behavior in the spatial fluctuations of the
velocity fields, characterized by bursts of activity.  Intermittency
is most commonly attributed to spatial fluctuations in the energy
dissipation rate, $\epsilon$ (note that $\epsilon$ was assumed
constant above) \cite{Frisch-Book-1995}.  Accounting for this
intermittency and the resultant adjustments to the Kolmogorov theory
is still an active area of research.

Intermittency is also a feature of velocity statistics in the
Lagrangian frame of three-dimensional turbulence.  Recent experimental
\cite{Xu-PRL-2006} and numerical \cite{Biferale-PRL-2004} measurements
of Lagrangian velocity difference statistics
demonstrate stronger deviations from similarity than were displayed in
the Eulerian statistics.  Though there is no exact result for
$D^{(n)}$ as there is for $S^{(n)}$, a dimensional argument suggests
that $D^{(n)}(\tau) \propto (\epsilon \tau)^{n/2}$ in the inertial
energy range of scales. A more intermittent Lagrangian signal is
perhaps not surprising; Kraichnan demonstrated
\cite{Kraichnan-PhF-1970} that even a non-intermittent Eulerian
velocity field (albeit an unphysical one) could produce intermittent
Lagrangian statistics.

As in three-dimensional turbulence, there are exact relations and
similarity results for Eulerian statistics in two-dimensional
turbulence
\cite{Paret-PhF-1998,Boffetta-PRE-2000,Tsang-PRE-2005,Eyink-PRL-1995}.
These results apply to both the inertial energy range of
two-dimensional turbulence and the inertial enstrophy range.  Unlike
three-dimensional turbulence, the Eulerian signal of the inertial
energy range of two-dimensional turbulence is non-intermittent and
self-similar, yielding scaling exponents for $S^{(n)}(r)$ in agreement
with the expected $r^{n/3}$ \cite{Boffetta-PRE-2000,Paret-PhF-1998}.
Similarly, scaling exponents of $S^{(n)}(r)$ for the Eulerian
enstrophy range have the expected $r^{n}$ behavior of the smooth
velocity field at those scales.  Although the scaling exponents in the
enstrophy range are self-similar, the velocity difference moment may
not be an adequate measurement tool to determine self-similarity for
such smooth fields.  This observation has resulted in the
investigation of ``inverse statistics'' \cite{Biferale-PRL-2001}.
Analysis of our data using the inverse-statistics approach will be
presented elsewhere.

Whereas Eulerian results have been measured previously for 2D
turbulence \cite{Paret-PhF-1998,Tsang-PRE-2005}, corresponding results
for Lagrangian velocity statistics have not been reported, perhaps
owing to the difficulty in producing a two-dimensional system with
significantly large scaling ranges for both energy and enstrophy.  In
three-dimensions, there is only a single inertial range in which the
energy cascade range lies between the outer scale and the viscous
scale.  In two-dimensions there are two such ranges: an inverse-energy
range analagous to the three dimensional energy range, and the direct
enstrophy range.  One therefore needs inertial behavior and
measurements over a larger range of scales in two-dimensional systems
than is needed in three.  Unfortunately, it is typical for
two-dimensional systems to have small ranges, which results in
ambiguous scaling and a lack of self-consistency in how the ranges are
defined.

When the ranges in question are small and the moments do not show
unambigous scaling behavior, no strong conclusions about scaling
exponents can be made.  Weaker conclusions can be drawn,
however, using techniques such as extended self-similarity (ESS)
\cite{Benzi-PRE-1993}, which explores the shape of the probability
distribution functions (PDFs) within a range of scales.  If the PDFs
exhibit a range over which their shape is either constant, or has a
particular trend, it is possible to extrapolate to the case where
one had a more extended range with a given scaling behavior.

In this paper we report measurements of velocity difference
statistics in both the Eulerian and Lagrangian frames of
two-dimensional turbulence.  Moreover, we report these results
for simultaneous energy and enstrophy ranges in 
our current system that is not significantly different from prior
experiments.  Owing to the limited extent
of the ranges, we do not attempt to directly extract exponents of
moments.  Instead, we present ESS analysis of the {\em local} scaling
exponents and demonstrate that the shape of the difference PDF's
evolves very differently in the Eulerian and Lagrangian frames.
If this shape evolution applies for larger ranges, we can conclude
that the velocity difference statistics in the Eulerian
frame for both the energy and enstrophy scaling ranges are
self-similar, whereas in the Lagrangian frame they are not.

\section{Experimental \label{sec: Experimental}}

Two-dimensional turbulence possessing simultaneous inverse energy and
direct enstrophy cascades must be continuously
forced \cite{Kraichnan-RPP-1980}.  One method for such forcing,
originally pioneered by Dolzhansky in 1979 \cite{Bondarenko-FAO-1979},
is to subject a current carrying fluid to external magnetic fields.
This method of forcing has since evolved into the stratified
electromagnetic layer\cite{Paret-PRL-1997} which has become a common
systems for the study of 2D turbulence.

\begin{figure}[b]
\includegraphics[width=3.25in]{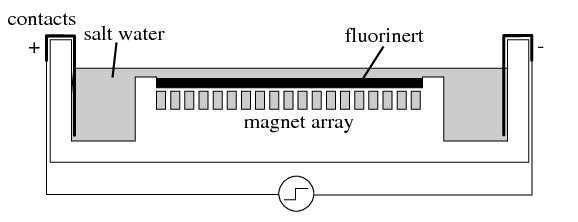}
\caption{\label{fig: Apparatus} Schematic illustration of the
  stratified-layer 2D turbulence experiment.}
\end{figure}
  
The stratified electromagnetic layer consists of a dense salt water
layer, typically $0.3$ cm deep and $20$ cm $\times$ $20$ cm square,
underneath a less dense solution of roughly the same depth.  A current
is passed in the plane of the layer, and the layers are subject to a
spatially varying magnetic field penetrating them vertically.  The
resultant Lorentz force drives fluid motion.  The evolution of the
stratified layer system is expected to approximate the forced/damped
2D Navier-Stokes equation,
\begin{equation}
\frac{\partial u_i}{\partial t} + u_s\frac{\partial u_i}{\partial x_s} = -
\frac{\partial p}{\partial x_i} + \nu \frac{\partial^2 u_i}{\partial x_i^2} -
\alpha u_i + F_i,
\label{eq: NavierStokes}
\end{equation}
supplemented by the incompressibility condition $\partial_i u_i = 0$,
where ${\bf u}$ is the fluid velocity field, $p$ is the density
normalized pressure, ${\bf F}$ is the external electromagnetic
forcing, $\nu$ is the fluid kinematic viscosity, and Einstein
summation is used throughout.  The linear term with coefficient
$\alpha$ represents the effects of frictional drag owing to the
container bottom. Although such a description is broadly consistent
with a number of experimental results, the detailed correspondence of
the electromagnetic layer flow with the linearly-damped 2D
Navier-Stokes equation has not been performed
\cite{Paret-PhF-1998,Rivera-PRL-2005}.

In the set of experiments presented here, the system described above
is modified slightly by replacing the lower layer of fluid with
Fluorinert FC-75 and the upper layer by a dense salt water solution of
$20$\% by mass NaCl with a small amount of liquid detergent added to
lower surface tension and help with dissolution of tracer
particles. The Fluorinert has a density $1.8$ times
that of water with nearly the same viscosity, which allows for much
stronger stratification than in the case of two salt-water
solutions. It is also a strong dielectric so that only the upper salt-water 
layer is electromagnetically forced. Finally, and perhaps most
importantly, the Fluorinert and water are immiscible.  These combined
features allow the Fluorinert system to maintain stratification
indefinitely, allowing the salt water layer to be driven harder than
in previous systems \cite{Paret-PRL-1997}.

\begin{figure}[t]
\includegraphics[width=3.5in]{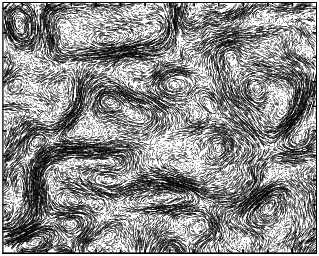}
\caption{\label{fig: RawTracks} Raw particle tracks obtained from the
  stratified layer over four consecutive frames ($0.07$ s).}
\end{figure}

The experimental apparatus, shown schematically in Fig.~\ref{fig:
  Apparatus}, consists of a $0.3$ cm thick layer of salt-solution
suspended over a $0.3$ cm thick layer of Fluorinert.  The layers are
contained in a $40$ cm $\times$ $20$ cm box with reservoirs at each
end across which copper electrodes are placed in the fluid. An
alternating square-wave current with frequency $0.5$ Hz and amplitude
$0.75$ amps is driven through the salt solution.  A set of $1.27$ cm
diameter rare-earth magnets of approximately $0.7$ T residual field
strength are arranged with alternating field direction in a $20$ cm
$\times$ $20$ cm square array with a period of $2.54$ cm and oriented
at $45^\circ$ with respect to the current direction.  The combination
of the current and the magnetic field produces a Kolmogorov-like
forcing \cite{Bondarenko-FAO-1979,Rivera-PRL-2005} of alternating shear
bands with the shear direction along ${\hat y}$ and periodicity
$r_{inj} \approx 1.8$ cm in the ${\hat x}$ direction, which implies
$k_{inj} \equiv 2 \pi/r_{inj} = 3.5$ rad/cm.  Using the layer depths
of $0.3$ cm yields $\alpha \approx 0.125$ Hz for the frictional
coupling assuming a simple linear shear in the Fluorinert.  The salt
solution upper layer has a viscosity around $1.15$ that of water.

To obtain Eulerian velocity fields and particle trajectory
information, the upper salt-solution layer was seeded with
polycrystalline powder with mean diameter $75$ $\mu m$ and density
$0.98$ gm/cc. Images of the particle fields, illuminated using several
Xenon short-arc flash lamps, were obtained with a $1280$ $\times$
$1024$ pixel CCD camera at a frame rate of $60$ Hz.  The velocity
field was obtained from image pairs using particle tracking
velocimetry derived from two earlier
methods\cite{Ishikawa-MST-2000,Ohmi-MST-2000}.  From a typical pair of
images, of order $3 \times 10^4$ particle tracks were obtained and
then interpolated to a $126 \times 100$ velocity field array.  A
typical field of raw particle tracks is shown in Fig.~\ref{fig:
  RawTracks}.

\begin{figure}[b]
\includegraphics[width=3.5in]{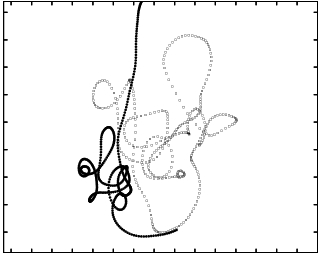}
\caption{\label{fig: Trajectories} Typical time-spliced real particle
  trajectory and a trajectory computed from dynamic velocity field
  data .}
\end{figure}

From this raw data one can obtain particle trajectories in two ways:
splice together raw particle tracks or use the interpolated Eulerian
fields to generate particle trajectories by solving the advection
equation.  The later technique will be used here. The generated
particle tracks solve the equation
\begin{equation}
\frac{d {\bf r}}{d t} = {\bf u}({\bf r}(t),t),
\end{equation}
where ${\bf r}(t)$ is the position of the tracer particle at time $t$,
bicubic interpolation is used to approximate ${\bf u}$ at the particle
position, and fourth-order Runga Kutta is used to perform the time
integration. Typical time-spliced particle trajectories and generated
trajectories are shown in Fig.~\ref{fig: Trajectories}.

The turbulent state that we obtain for the stratified layer system
described above can be characterized by several of its Eulerian
ensemble-averaged statistics.  The average energy per unit mass $E =
u_{rms}^2/2$ is $8.4$ cm$^2/$s$^2$ ($u_{rms}=4.1$ cm/s) and the
average enstrophy $\Omega = \omega_{rms}^2/2$, where the vorticity
$\omega \equiv {\bf \nabla} \times {\bf u}$, is $51$ s$^{-2}$
($w_{rms}=10.1$ s$^{-1}$).

\section{Results and Discussion\label{sec: ResultsAndDiscussion}}

\begin{figure*}
\includegraphics[width=3.5in]{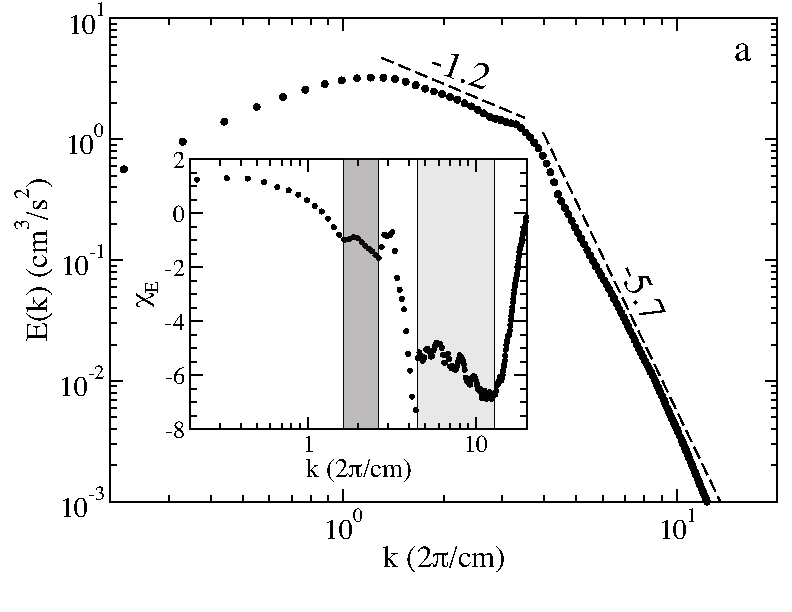}
\includegraphics[width=3.5in]{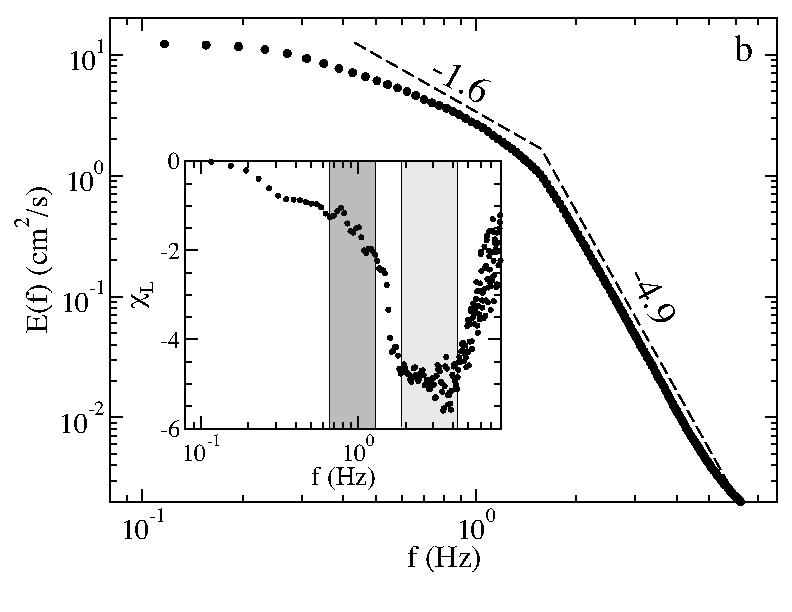}
\caption{\label{fig: Spectra} (a) Eulerian and (b) Lagrangian
  energy spectra.  Insets show the behavior of the respective
  spectra's local scaling exponent.  The proposed ranges of wave
  numbers (Eulerian) and frequencies (Lagrangian) over which there is
  an inverse-energy range is denoted by a dark bar in the insets.  The
  proposed direct-enstrophy range is likewise indicated by a light
  bar.  Best fit power laws over the suggested ranges are displayed by
  lines with the scaling exponents given.  See the text for the exact
  ranges and power laws.}
\end{figure*}

Our system displays two different ranges corresponding to
predominantly inverse-energy transfer \cite{Chen-PRL-2006} for scales
larger than the injection scale and direct-enstrophy transfer for
scales smaller than the injection scale.  Because of the relatively
limited range of scales available in the experiments, it is important
to firmly establish the extent of the spatial and temporal scales for
each transfer range.  Empirically, this is best done utilizing
spectra: we find that turbulent range transitions are sharper in
Fourier space than in real space.  The Eulerian and Lagrangian energy
spectrum for data obtained from the electromagnetic cell are shown in
Figs.~\ref{fig: Spectra}a and b, respectively.  For the Eulerian
spectrum, $E(k)$ is calculated for individual velocity fields using
$\tilde{u}_i(k)\tilde{u}_i^*(k)/2$ and averaged over velocity fields
from the entire run.  The Lagrangian spectrum, $E(f)$, is calculated
from the Lagrangian velocity correlation $\langle u_i(t)u_i(t+\tau)
\rangle$ using the Wiener-Kinchin theorem.

We begin by identifying the length scales over which there is an
inverse-energy range.  According to previous results
\cite{Kraichnan-RPP-1980,Paret-PhF-1998,Boffetta-PRE-2000}, this range
occurs at wave numbers smaller than the injection wave number
$k_{inj}=2\pi/r_{inj}=3.5$ rad/cm and is characterized by a spectral
scaling of $E(k) \propto k^{-5/3}$.  To identify this range, we plot
in the inset the local spectral exponent,
\begin{eqnarray}
\chi_E \equiv \frac{d\{log(E(k)/E(k_{inj}))\}}{d\{log(k/k_{inj})\}}. 
\end{eqnarray}
The value $-5/3$ does not occur over any extended range.  There is,
however, a tight band of wave numbers smaller than $k_{inj}$ over which
the values of $\chi_E$ are close to $-5/3$, namely $1.65$ cm$^{-1}$ $<
k < 2.64$ cm$^{-1}$ (the dark bar in the inset plot).  At wave numbers
smaller than this range the spectral slope rises through zero and
becomes positive corresponding to the large-scale saturation peak.
wave numbers just in excess of this range are associated with the
injection scales.  We take this tight band as the inverse-energy
range; the choice of this tight band as a ``range'' will be justified
in later analysis.  A best fit power law to the range yields a
spectral exponent of $-1.2$.  Taking the point at $k=2.64$ cm$^{-1}$
and assuming $E(k)=C^{E}_0 \epsilon^{2/3} k^{-5/3}$, where $\epsilon$
is the scale-to-scale energy transport rate (roughly the large-scale
dissipation $\alpha u^2_{rms}$), yields a value for the
Eulerian-Kolmogorov constant of $C^{E}_0 \approx 6$, in agreement with
earlier numerical and experimental work
\cite{Paret-PhF-1998,Boffetta-PRE-2000}.

The same procedure is used to establish the direct-enstrophy range at
wave numbers greater than $k_{inj}$.  The spectra in the
direct-enstrophy range is expected to scale as $E(k) \propto k^{-3}$
in the absence of frictional dissipation \cite{Kraichnan-RPP-1980}.
With frictional dissipation, as is the case in the electromagnetic
cell, the scaling exponent is expected to become more negative
\cite{Tsang-PRE-2005}.  For wave numbers $4.5$ cm$^{-1} < k < 13$
cm$^{-1}$, indicated by a light bar in the inset, $\chi$ has an
average value of $-5.7$.  This range is taken to be the
direct-enstrophy range. Note that the scaling exponent is extremely
steep when compared to the dissipation free expectation of $-3$,
indicative of a large amount of external dissipation.  The upper-limit
of the direct-enstrophy range $k=13$ cm$^{-1}$ corresponds to a length
scale of $5$ mm, which is comparable to the fluid layer depth of $3$
mm where one might expect 3D effects to become important.

Establishing the time scales of the inverse-energy and
direct-enstrophy ranges in the Lagrangian frame is done in the same
manner as for Eulerian statistics.  The dimensional expectation for
the energy range is that $E(f) \propto f^{-2}$ for frequencies smaller
than the energy injection frequency, $f_{inj} = 1.66$ Hz, which is
about the eddy rotation frequency of an injection scale vortex.  As in
the Eulerian spectra, the Lagrangian spectra do not achieve a range
over which the spectral scaling, measured by the exponent $\chi_L$
defined in the same way as $\chi_E$ and shown in the inset, achieves a
constant $-2$ value. It also does not have as clearly defined a range
as the Eulerian spectrum. The region marked by the dark bar
corresponding to frequencies between $0.66$ Hz and $1.28$ Hz,
characterized by a spectral slope increase from a $-2$ value to about
$-1$, should be the inverse-energy-range. Below this range the
spectral exponent is less than $1$ corresponding to large-scale
saturation, and above is the abrupt spectral exponent change
associated with $f_{inj}$.  A best fit power law to this range yields
an exponent of $-1.6$.  Taking the point at $f=1.28$ Hz and assuming
$E(f)=C^{L}_0 \epsilon f^{-2}$ yields a Lagrangian-Kolmogorov constant
of $C_0^L \approx 10$.  The enstrophy range, for which a simple
dimensional prediction does not exist, is taken to be the range of
frequencies spanning $1.87$ Hz to $4.27$ Hz, over which the spectral
exponent is around $-4.9$.  The frequency $4.27$ Hz corresponds to a
$0.23$ s time scale, which is the same as the Lagrangian correlation
time calculated by integration of the normalized two-point
correlation.  The steep spectral exponent is in agreement with earlier
geophysical observations \cite{Rupolo-JPO-1996}.

We now consider Eulerian and Lagrangian velocity statistics as
measured by $n^{\rm th}$ order moments of velocity differences, {\it
  i.e.}, structure functions, $S^{(n)}(r)$ and $D^{(n)}(\tau)$,
respectively.  Averages for the structure function calculation are
taken over ensembles of realizations, and spatial and temporal
homogeneity of signals is assumed so that the moments do not depend on
absolute position ${\bf x}$ or time $t$.  For the inverse energy range
of two-dimensional turbulence, we have dimensional predictions for
scaling exponents: $S^{(n)}(r) \propto (\epsilon r)^{n/3}$ and
$D^{(n)}(\tau) \propto (\epsilon \tau)^{n/2}$.  In the enstrophy range
for Eulerian statistics, we expect $S^{(n)}(r) \propto r^{n}$,
indicating a smooth velocity field for which the linear order term in
the Taylor expansion is dominant.  For Lagrangian statistics, at small
time scales where acceleration is approximately constant, we expect
$D^{(n)}(\tau) \propto \tau^{n}$. Thus, we expect scaling in the
enstrophy range to have exponents between $n$ and $n/2$.

\begin{figure*}
\includegraphics[width=3.5in]{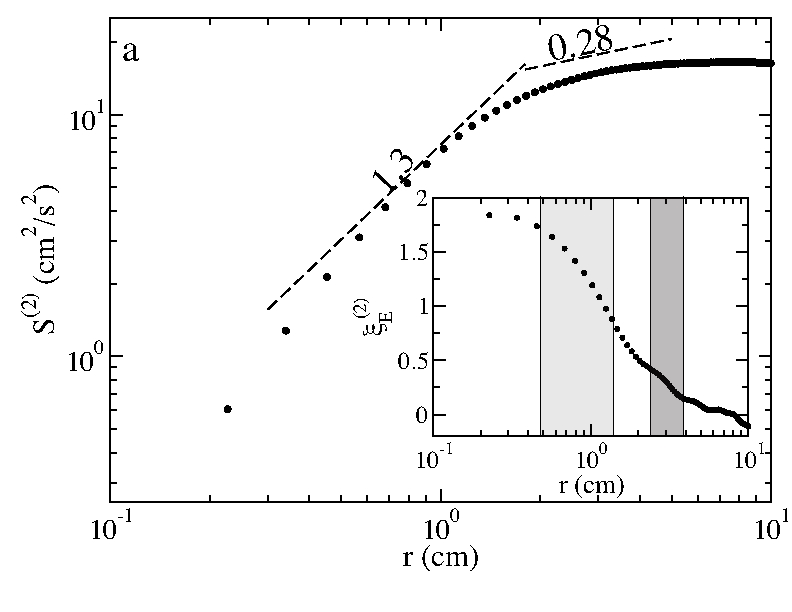}
\includegraphics[width=3.5in]{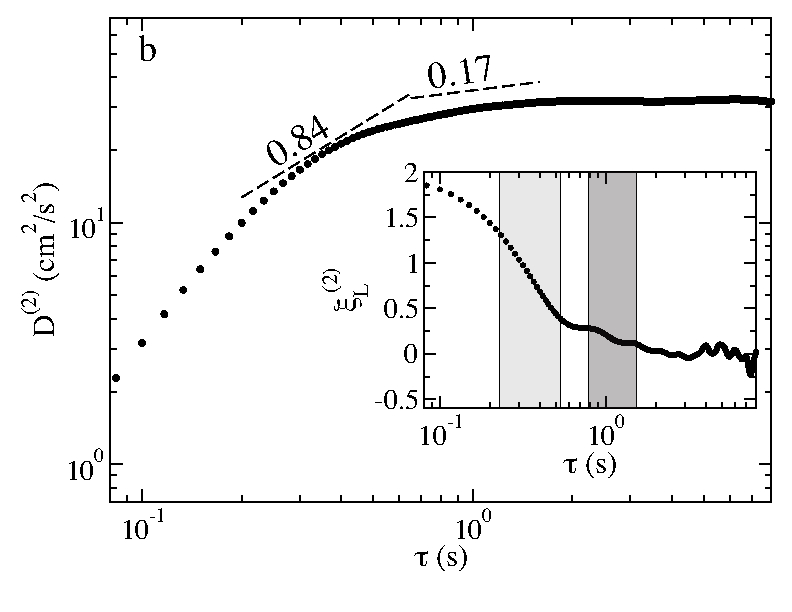}
\caption{\label{fig: SecondMoments} (a) Eulerian and (b) Lagrangian
  second-order structure functions.  Insets show the behavior of the
  respective structure function local scaling exponents.  The ranges
  of length (Eulerian) and time (Lagrangian) for the inverse-energy
  regime, determined from spectra, are denoted by dark bars in the
  insets.  The direct-enstrophy ranges, likewise determined from
  spectra, are indicated by light bars.  Best fit power laws over the
  suggested ranges are displayed by lines with the scaling exponents
  given.  See the text for the exact ranges and power laws.}
\end{figure*}

We begin by looking at the second moments (n=2) which are displayed in
Fig.~\ref{fig: SecondMoments}.  Inset in these plots are the
local exponents for the second moments, defined as
\begin{eqnarray}
\xi_E^{(n)} \equiv \frac{d\{log(S^{(n)}/u^n_{rms})\}}{d\{ log(r/r_{inj})  \}},\\
\xi_L^{(n)} \equiv \frac{d\{log(D^{(n)}/u^n_{rms})\}}{d\{ log(\tau/\tau_{inj})  \}}.
\end{eqnarray}
A scaling range where $\xi$ has a constant value is not observed.
This is unsurprising given the small extent of the ranges as
identified by spectral analysis.

The second moments are frequently used in lieu of spectral analysis to
identify scaling ranges.  This is done because, in the limit of long
scaling ranges, the results should yield similar ranges (the
Wiener-Kinchen theorem links the spectra with the second-moments
through the correlation function ).  For small ranges, however, simply
assuming this correspondence is dangerous.  If we determine temporal
and spatial scales utilizing these moments by, for example, bracketing
ranges with the expected scaling exponents, we get inconsistent
results.  In particular, we expect the Eulerian second moment to scale
as $r^{2/3}$ in the inverse energy range.  Looking for the region
where $\xi^{(2)}_E$ assumes a value of $0.66$ would yield a length
scale of around $1.44$ cm, which is significantly smaller than the
injection scale of $1.8$ cm, {\em i.e., in the enstrophy range of
  length scales}.  A similar phenomenon happens for time scales of the
Lagrangian second moment.  The time scale at which $\xi^{(2)}_L$
assumes the expected value of $1$ is well within the spectrally
identified enstrophy range.

Given the lack of correspondence between real and spectral space
results, it is interesting to ask if ranges identified spectrally are
reflected in their real space counterparts.  It is also valid to
consider why we use the spectral method for identifying ranges in the
first place since it was an empirical decision to do so. In the rest
of this paper, we demonstrate that although the spectral ranges are
not easily identified in the direct moments, they correspond directly
to features of the PDF shape, that is, when we investigate velocity
statistics using ESS.  For the moment, simply assuming the spectrally
identified ranges are correct (marked by a dark and light bar again),
we get best fit power laws to the second moments that significantly
deviate from expected behavior, see Fig.~\ref{fig: SecondMoments}.

\begin{figure*}
\includegraphics[width=3.5in]{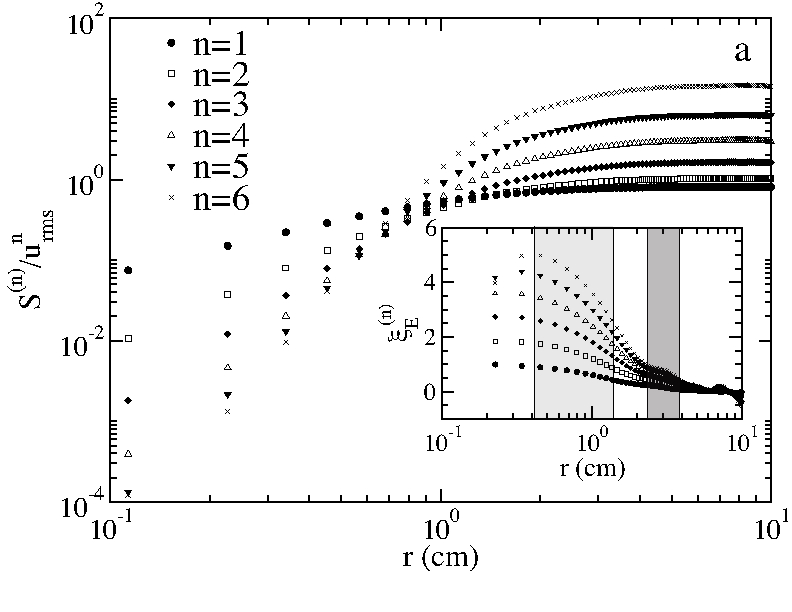}
\includegraphics[width=3.5in]{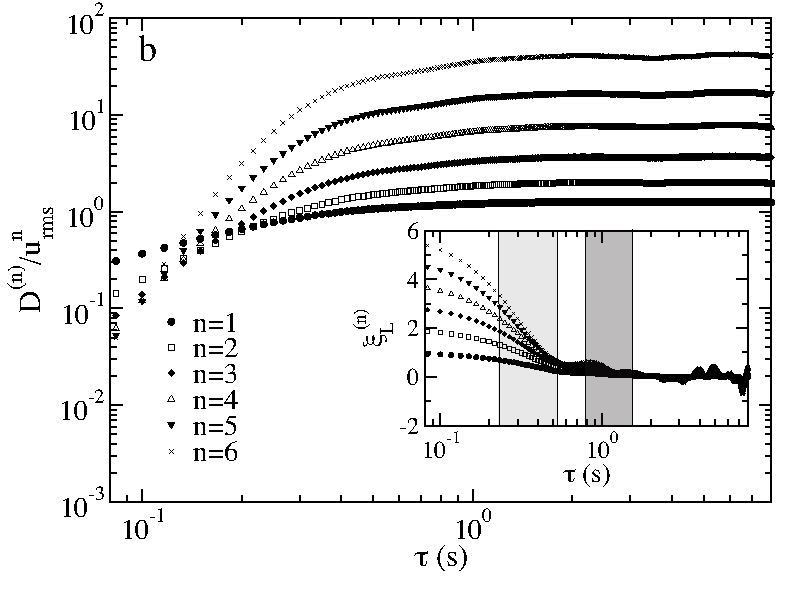}
\caption{\label{fig: HighMoments} (a) Eulerian and (b) Lagrangian
  n$^{\rm th}$-order structure functions normalized by $u_{rms}^n$.  Insets
  show the behavior of the respective structure function local scaling
  exponents.  The ranges of length (Eulerian) and time (Lagrangian)
  for the inverse energy regime, determined from spectra, are denoted
  by dark bars in the insets.  The enstrophy-ranges, likewise
  determined from spectra, are indicated by light bars.}
\end{figure*}

The Eulerian and Lagrangian velocity-difference moments of order $1
\leq n \leq 6$ are shown in Fig.~\ref{fig: HighMoments}, and the
insets show the local exponents with dark and light bars to indicate
the spectrally-identified scaling ranges.  To fit exponents
confidently to high-order moments, scaling is needed over an extensive
range.  Unsurprisingly, based on the previous spectral analysis, no
such ranges are visible at moments of any order.  Although long ranges
are not seen for either the Eulerian or Lagrangian data, the exponents
of the moments behave in a very similar manner at different order,
suggesting distinct changes in the behavior of the local exponents
over the spectrally identified ranges. Because constant exponents are
not observed in the measured range, we are forced to draw conclusions
based on ESS analysis.

\begin{figure*}
\includegraphics[width=3.5in]{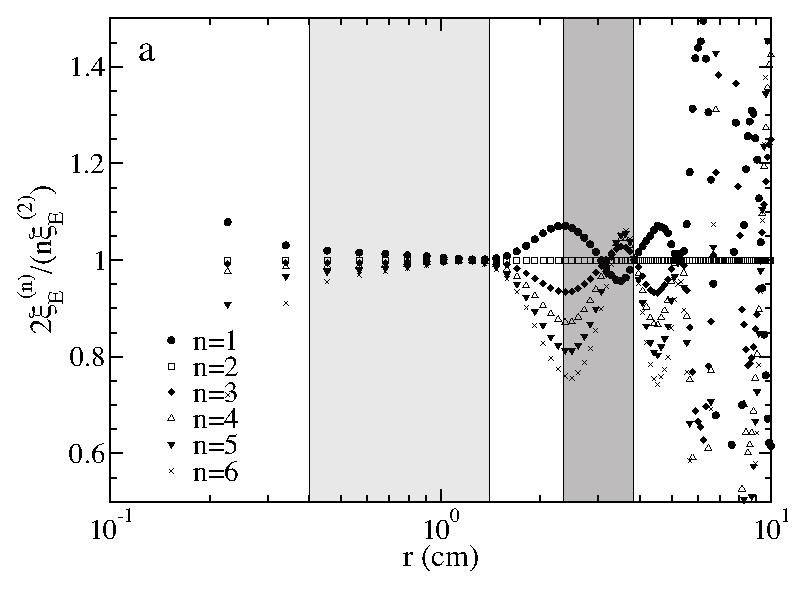}
\includegraphics[width=3.5in]{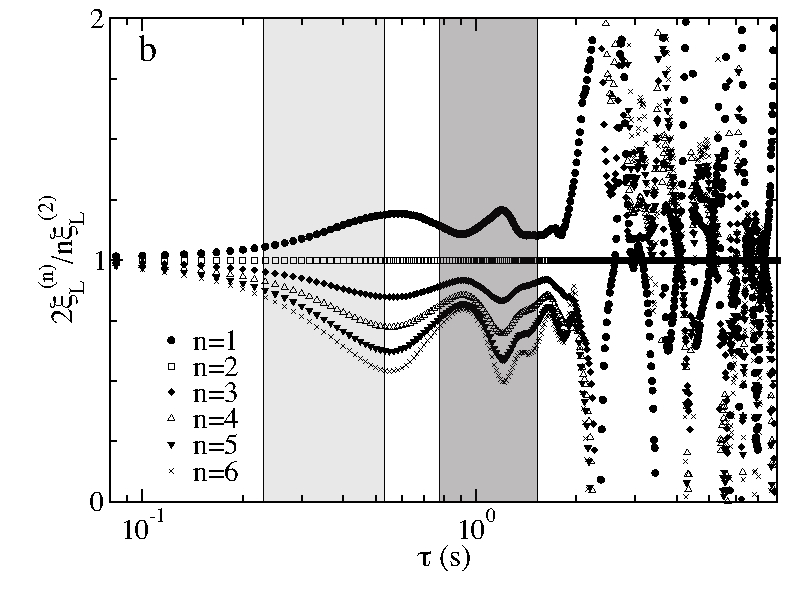}
\caption{\label{fig: MomentScaling} The structure functions local
  exponent for orders $1 \leq n \leq 6$ normalized by $n\xi{(2)}/2$
  for (a) the Eulerian frame and (b) the Lagrangian frame. The ranges
  of length (Eulerian) and time (Lagrangian) for the inverse energy
  regime, determined from spectra, are denoted by dark bars in the
  insets.  The enstrophy-ranges, likewise determined from spectra, are
  indicated by light bars.  A range with value of unity for {\em all}
  of the normalized local exponent indicates that the PDF shape is
  constant.}
\end{figure*}

We normalize the local exponents by the second-order local exponent
and scale by the expected value of $n/2$, {\em i.e.}, we use ESS.  By
doing this normalization, we are looking at the change in PDF shape
over a range of scales.  This is shown in Fig.~\ref{fig:
  MomentScaling}.  There is very different behavior for the Lagrangian
and Eulerian analysis.  For self-similarity to hold over a range (that
is, an unchanging PDF shape), the value of the normalized local
exponent must be unity, for all $n$, over that range, {\it i.e.}, the
higher moments are simply scaling as a power of the second moment.
Given these observations, it is possible to draw some stronger
conclusions.

\begin{figure}
\includegraphics[width=3.5in]{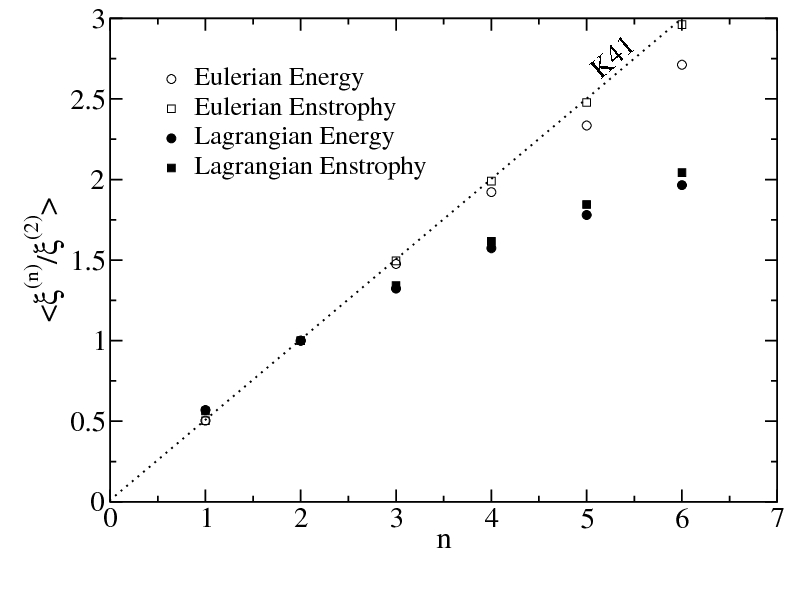}
\caption{\label{fig: Exponents} The normalized local exponents for the
  velocity difference moments averaged over the inverse-energy and
  direct enstrophy ranges in both the Eulerian and Lagrangian frame.
  Strict self-similarity over a given range would be indicated by
  collapse of the moments on the K41 (Kolmogrov 1941) line.}
\end{figure}

The spectrally-identified ranges, which seemed arbitrary and did not
clearly correspond to any particular behavior in the real-space
statistics, become more transparent.  The enstrophy range in the
Eulerian statistics is characterized by a near unity grouping of all
the moments.  At the upper-end of the Eulerian enstrophy range we see
a sharp change in the behavior of the exponents where they begin to
deviate from one.  This deviation reaches a peak at the low-end of the
inverse energy range and quickly collapses back to near unity.  The
conclusion we draw from this behavior is that, in our system, both the
energy and enstrophy range Eulerian velocity difference statistics are
tending to behave self-similarly ({\em i.e.}, normalized exponents
tending to unity). This is in agreement with earlier observations of
the unnormalized quantities.
\cite{Paret-PhF-1998,Boffetta-PRE-2000,Tsang-PRE-2005}.  We speculate,
given previous numerical work with more extensive ranges, that the
small deviation from unity that we observe, in the limit of a long
range well removed from injection effects, should settle down to
unity. The average value of the exponent ratio for both the Eulerian
energy and enstrophy ranges for a range of $n$ is displayed in
Fig.~\ref{fig: Exponents} as open symbols.  A line marked K41 (for
Kolmogorov 1941) indicates self-similar scaling.

The story is quite different for the Lagrangian results.  The spectral
ranges still find support, but in almost exactly the opposite manner:
divergences from unity grow in the energy and enstrophy ranges while
being reduced in the intermediate (injection) range (note that the
vertical scales are different between Fig.~\ref{fig: MomentScaling}a
and b).  Indeed, for the energy range, the maximal deviation from
unity occurs in the center of the range, collapsing back to unity only
after the outer scale is reached.  This allows us to speculate that,
even in the limit of long ranges, the energy range will continue to
deviate from unity and display a lack of similarity.  Interpretation
of what occurs in the enstrophy range is somewhat more complicated.
There is a deviation from unity, but whether this is a residual result
of a small range and the close proximity to the injection scale is
uncertain.  In the limit of a long range, the majority of the
enstrophy range may approach unity.  This would be unsurprising
because the spectral statistics indicate a very smooth enstrophy range
in the Lagrangian frame.  Therefore, we expect velocity difference
statistics to be dominated by first-order terms in the series
expansion as in the Eulerian case.  Without further measurements it is
difficult to extrapolate based on this data.  As with the Eulerian
data, the average local exponent over the energy and enstrophy ranges
for a range of $n$ is displayed in Fig.~\ref{fig: Exponents} as close
symbols.  Deviation from similarity is readily apparent when compared
with the Eulerian statistics.

The data at very large scales and times are not well enough converged
to draw any conclusions about statistical behavior past the outer
scale. The dissipative ranges (small $r$ and $\tau$), however, are
well converged and observably different.  The Eulerian range is
characterized by an increasingly non-self-similar dissipation range,
whereas the Lagrangian frame quickly collapses to a state of
similarity. For the Eulerian frame, this is most likely indicative of
a spatially inhomogenous viscous dissipation - a similar observation
is found in three-dimensional turbulence \cite{Frisch-Book-1995}.  For
the Lagrangian frame, the collapse to similarity is surprising.  It
may be misleading, however.  At these short times, linear order terms
in a expansion are dominant, and therefore velocity difference may not
be able to capture deviations from similarity well, much as is the
case for the velocity difference statistics in the Eulerian enstrophy
range.

\section{Conclusions}

We have considered velocity difference statistics in two-dimensional
turbulence for both the Eulerian and Lagrangian frames.  Eulerian
statistics are investigated and compared with earlier results, and
novel Lagrangian statistics are presented.  As in prior experiments,
the extent of the inverse-energy range and direct-enstrophy range are
limited.  The limited range has two major effects: the scaling
exponents are different in magnitude from the dimensional predictions,
and there is no easily identifiable correspondence between ranges in
spectra and behavior of moments.  These observations are not
unexpected for systems without extended scaling ranges.

Further analysis using ESS, however, yields a number of important
insights.  First, the results in the Eulerian frame are consistent
with the broadly self-similar results found in prior experiments and
simulations for velocty difference statistics, in spite of a lack of
scaling.  Second, the Lagrangian frame velocity difference statistics
are not self-similar, a conclusion more strongly evident in the
inverse-energy range than in the direct-enstrophy range.  Finally, the
dissipation ranges are not self-similar in the Eulerian frame, whereas
they collapse to self-similar form for the Lagrangian frame.

In addition to our main conclusions, we can provide two further
insights from an analysis perspective.  First is that careful
identification of spectral ranges are echoed in the behavior of the
moments local exponent ratios fairly precisely.  This was somewhat
surprising since it has been believed that moment ranges do not
necessarily line up in detail with spectral ranges, in particular when
those ranges are small.  The data here indicate otherwise.  Second,
the local exponent ratio is effectively a derivative quantity. It
measures the relative change in the behavior of the PDF's shape over a
range of scales.  Even when the moment ranges are small, the shape of
the corresponding PDF changes rapidly.

\section{Acknowledgments}

This work was performed for the U.S.~Department of Energy under
Contract \# DE-AC52-06NA25396.  The authors benefited from
conversations with Colm Connaughton and Mahesh Bandi.

\bibliography{VelStats}

\end{document}